\documentclass[prd,nofootinbib,preprint,superscriptaddress
]{revtex4}
 
\usepackage{graphicx}
\usepackage{amsfonts}

\newcommand{\be}{\begin{equation}}
\newcommand{\ee}{\end{equation}}
\newcommand{\bea}{\begin{eqnarray}}
\newcommand{\eea}{\end{eqnarray}}
\newcommand{\beq}{\begin{equation}}
\newcommand{\eeq}{\end{equation}}
\newcommand{\beqa}{\begin{eqnarray}}
\newcommand{\eeqa}{\end{eqnarray}}

\def\lsim{\lesssim}
\def\gsim{\gtrsim}

\begin{document}

\vspace*{.0cm}

\title{Probing Minimal Flavor Violation at the LHC}

\author{Yuval Grossman}\email{yuvalg@physics.technion.ac.il}
\affiliation{Department of Physics, Technion-Israel
  Institute of Technology, Technion City, Haifa 32000,
  Israel}
\author{Yosef Nir\footnote{The Amos de-Shalit chair of theoretical
    physics}}\email{yosef.nir@weizmann.ac.il} 
\affiliation{Department of Particle Physics,
  Weizmann Institute of Science, Rehovot 76100, Israel}
\author{Jesse Thaler}\email{jthaler@jthaler.net}
\affiliation{Department of Physics, 
University of California, Berkeley, CA 94720}
\affiliation{Theoretical Physics Group, Lawrence Berkeley National Laboratory, 
 Berkeley, CA 94720}
\author{Tomer Volansky}\email{tomer.volansky@weizmann.ac.il}
\affiliation{Department of Particle Physics,
  Weizmann Institute of Science, Rehovot 76100, Israel}
\author{Jure Zupan}\email{jure.zupan@fmf.uni-lj.si}
\affiliation{Department of Physics, University of Ljubljana, Jadranska 19, 1000
Ljubljana, Slovenia}
\affiliation{J. Stefan Institute, Jamova 39,
P.O. Box 3000, 1001 Ljubljana, Slovenia\vspace*{1cm}}

\vspace*{1cm}
\begin{abstract}
  If the LHC experiments discover new particles that couple to the
  Standard Model fermions, then measurements by ATLAS and CMS can
  contribute to our understanding of the flavor puzzles. We
  demonstrate this statement by investigating a scenario where extra
  SU(2)-singlet down-type quarks are within the LHC reach. By
  measuring masses, production cross sections and relative decay rates,
  minimal flavor violation (MFV) can in principle be excluded.
  Conversely, these measurements can probe the way in which MFV
  applies to the new degrees of freedom. Many of our conclusions are
  valid in a much more general context than this specific extension of
  the Standard Model. 
\end{abstract}

\maketitle

\section{Introduction}
Significant progress in our knowledge of flavor physics has been
achieved in recent years. The main contributions have come from the
two B factories, Belle and BaBar, and from the Tevatron detectors, CDF
and D0~\cite{Yao:2006px,Group:2007cr}. The absence of evidence in the
data for new physics poses a puzzle: there are good reasons to think
that there is new physics at or below the TeV scale, yet any effective
non-renormalizable flavor-changing operators suppressed by the TeV
scale must mimic the Standard Model (SM) flavor structure with
excellent accuracy, reaching in some cases a level of one part in
$10^6$. The question of how and why that happens is often referred to as
the ``new physics flavor puzzle.''

We will soon enter a new era in high energy physics---the LHC era.
The LHC experiments should first answer the crucial question of
whether there is indeed new physics at the TeV scale, as suggested by
the hierarchy problem and weakly-interacting dark matter proposals. If
the answer is in the affirmative, then the LHC also offers new
opportunities in exploring the new physics flavor puzzle. If new
particles that couple to SM fermions are discovered,
then measurements of their spectrum and of their couplings will help
elucidate the basic mechanism that has so far screened the flavor
effects of new physics. The main goal of this work is to demonstrate
how high-$p_T$ processes, measured by ATLAS and CMS, can shed light on
flavor issues.

Of course, the implications of new physics on flavor are highly
model-dependent.  At energies much below the electroweak scale, the
flavor effects of new physics can be entirely captured by a series of
higher-dimension operators, but at LHC energies, flavor-changing
processes can occur via the production and decay of new on-shell
particles.  In models like supersymmetry (SUSY) with numerous new
partners and the potential for long cascade decays, flavor questions
can in principle be addressed \cite{Hinchliffe:2000np}, but in the
quark sector this is only possible after disentangling many
model-dependent effects like gaugino-Higgsino mixing angles and the
mass ordering of left- vs. right-handed squarks.  For purposes of
studying how flavor might be probed at the LHC, it is therefore
desirable to analyze models (which might be one sector of a more
complete theory) for which flavor has an unambiguous effect on LHC
signatures.

A simple and rather generic principle that can guarantee that low
energy flavor changing processes would show no deviations from SM
predictions is that of {\it minimal flavor violation}
(MFV)~\cite{D'Ambrosio:2002ex,Buras:2000dm,Buras:2003jf}. The basic
idea can be described as follows (a more rigorous definition is given
in the next section). The gauge interactions of the SM are universal
in flavor space. The only breaking of this flavor universality comes
from the three Yukawa matrices, $Y_U$, $Y_D$ and $Y_E$. If this
remains true in the presence of new physics---namely $Y_U$, $Y_D$ and
$Y_E$ are the only flavor non-universal parameters---then the model
belongs to the MFV class. We use the concrete question of whether
ATLAS and CMS can {\it test} the principle of MFV in order to explore
the flavor physics potential of these experiments.

To do so, we further choose a specific example of new physics. We
augment the SM with down-type, vector-like heavy fermions, $B_L$ and
$B_R$, that transform as $(3,1)_{-1/3}$ under the SM gauge group (for
a review see, for example, \cite{Branco:1999fs}).  To be relevant to
our study, at least some of the new quarks must be within the reach of
the LHC, and they must couple to the SM quarks. We assume that MFV
applies to this extension of the SM, and we ask the following
questions:
\begin{itemize}
\item What are the possible spectra of the new quarks?
\item What are the possible flavor structures of their couplings to
  the SM quarks?
\item Can the LHC exclude MFV by measurements related to these quarks?
\item In case that MFV is not excluded, can the LHC be used to support
MFV?
\end{itemize}
While in this study we concentrate only on a specific representation of
the extra quarks, many of the lessons that we draw have a much more
general applicability beyond our specific example.

In section \ref{theory} we introduce the notion of minimal flavor
violation and its consequences for a SM extension with extra
vector-like down-type quarks. The resulting spectrum and decay patterns are
discussed in section \ref{sec:spe}. In section \ref{sec:lhc} we
examine how experiments at LHC can refute or give support to the MFV 
hypothesis, and then summarize our conclusions in section
\ref{sec:con}.

\section{The theoretical framework}\label{theory}
The SM with vanishing Yukawa couplings has a large global
symmetry, $U(3)^5$. In this work we concentrate only on the quarks.
The non-Abelian part of the flavor symmetry for the quarks can be
decomposed as follows:
\begin{eqnarray}
  \label{eq:7}
  G_{\rm Flavor}= SU(3)_Q \otimes SU(3)_D\otimes SU(3)_U.
\end{eqnarray}
The Yukawa interactions ($H_c=i\tau_2 H^*$),
\begin{eqnarray}
  \label{eq:lagy}
{\cal L}_Y=\overline{Q_L}Y_D D_R H + \overline{Q_L}Y_U U_R H_c ,
\end{eqnarray}
break the $G_{\rm Flavor}$ symmetry. The Yukawa couplings can thus
be thought of as spurions with the following transformation properties
under $G_{\rm Flavor}$:
\begin{eqnarray}
  \label{eq:10}
  Y_D\sim(3,\bar3,1),\qquad Y_U\sim(3,1,\bar3).
\end{eqnarray}

We extend the SM by adding vector-like quarks $B_L$ and $B_R$ of
electric charge
$-1/3$. In general, extending the SM with the $B_L$ and $B_R$ fields gives
three new types of Yukawa and mass terms:
\begin{eqnarray}
  \label{eq:lagb}
{\cal L}_B=\frac{m_2}{v}\overline{Q_L}Y_B B_R H
+M_1\overline{B_L}X_{BD}D_R+M_2\overline{B_L}X_{BB}B_R.
\end{eqnarray}
Our assumption is that the mass parameters $M_1$ and $M_2$ are much
larger than the weak scale, while $m_2$ is of order the weak scale.
If the three new matrices $Y_B$, $X_{BD}$ and $X_{BB}$ had a generic
flavor structure, unrelated to that of $Y_D$ and $Y_U$, the deviations
from the SM predictions for flavor changing processes would exclude
the model, unless the mass scale for the new quarks is very high, well
beyond the LHC reach
\cite{Aguilar-Saavedra:2002kr,Andre:2003wc,Yanir:2002cq}.  We thus
impose the criterion of minimal flavor violation (MFV): all the
Lagrangian terms constructed from the SM fields, the $B_L$ and $B_R$
fields, and $Y_{D,U}$, must be (formally) invariant under the flavor
group $G_{\rm Flavor}$.

We are interested in the case that the new quarks couple to the SM
ones at renormalizable level. Then, we are led to models where the
$B_L$ and $B_R$ fields cannot be singlets of $G_{\rm Flavor}$.  (In
fact, the same result follows from the demand that the new fields have
allowed decays into SM fields.)  This is a general result: MFV (and
the requirement of coupling to SM fields) implies that the number of
extra vector-like quarks is at least three. Since there are many
options for $G_{\rm Flavor}$ charge assigments, for concreteness we
further narrow our scope to the cases where $B_L$ and $B_R$ are
singlets of $SU(3)_U$ and transform as $(3,1)$ or $(1,3)$ under
$SU(3)_Q \otimes SU(3)_D$. There are four possible combinations of
flavor-charge assignments to the $B_{L,R}$ fields.  These assignments
are given in Table \ref{tab:flarep}.

\begin{table}[t]
\begin{center}
\begin{tabular}{l|ccc|ccc} \hline\hline
\rule{0pt}{1.2em}%
Model & Quark field & $SU(3)_Q$ & $SU(3)_D$ & $Y_B$ & $X_{BB}$ & 
$X_{BD}$ \cr \hline\hline
& $Q_L$ & $3$ & $1$ & & &  \cr
& $D_R$ & $1$ & $3$ & & &  \cr
& $Y_D$ & $3$ & $\bar3$ & & & \cr
& $Y_UY_U^\dagger$ & $1+8$ & $1$ & & & \cr
\hline\hline
\textbf{QD} & $B_L$ & $3$ & $1$ & & & \cr
  & $B_R$ & $1$ & $3$ & $D_3^m Y_D$ & $D_3^M Y_D$ & $0$ \cr\hline
\textbf{DD} & $B_L$ & $1$ & $3$ & & & \cr
   & $B_R$ & $1$ & $3$ & $D_3 Y_D$ & $1$ & $0$ \cr \hline
\textbf{DQ} & $B_L$ & $1$ & $3$ & & & \cr
    & $B_R$ & $3$ & $1$ & $D_3^m$ & $Y_D^\dagger D_3^M$ & ($0$) \cr\hline
\textbf{QQ} & $B_L$ & $3$ & $1$ & & & \cr
   & $B_R$ & $3$ & $1$ & $D_3^m$ & $D_3^M$ & $D_3^Y Y_D$ \cr \hline\hline
\end{tabular}
\end{center}
\caption{The possible flavor assignments for vector-like quarks that
  transform as $(3,1)_{-1/3}$  under the SM gauge group. Here, we
  assume that $B_L$ and $B_R$ transform either as $(1,3)$ or $(3,1)$
  under $SU(3)_Q \times SU(3)_D$. The model names are determined in a
  self-evident way from the flavor assignments. The last three columns
  give the flavor structure for the new Lagrangian terms in
  Eq.~(\ref{eq:lagb}), assuming MFV. The matrices $D_3 \sim {\rm
  diag}(1,1,1+d_3)$ parametrize the breaking of $SU(3)_Q$ by the top
  Yukawa. In models \textbf{QD} and  \textbf{DD}, $X_{BD}$ can be
  taken to be zero by a $D_R - B_R$ rotation. The ``$(0)$'' in model \textbf{DQ} indicates a value that must be fine-tuned to get the right SM quark spectrum.}\label{tab:flarep}
\end{table}

Once the $G_{\rm Flavor}$-representations of the new fields are
defined, the flavor structure of their couplings in
Eq.~(\ref{eq:lagb}) is determined. The flavor structures are also
given in Table \ref{tab:flarep}. For the examples we are considering,
there are only two relevant spurions, $Y_D$ and $Y_UY_U^\dagger$.
Without loss of generality, we work in a basis where $Y_U$ is
diagonal. To a good approximation we can neglect the Yukawa couplings
of the up and charm quarks, and take $Y_UY_U^\dagger\sim {\rm
  diag}(0,0,1)$. The effect of $Y_UY_U^\dagger$ can be captured by
the combination
\begin{eqnarray}
  \label{eq:12}
  D_3 \equiv {\bf 1}+ d_3 Y_UY_U^\dagger\sim{\rm diag}(1,1,1+d_3),
\end{eqnarray}
where ${\bf 1}$ is the $3\times3$ unit matrix and $d_3={\cal O}(1)$.
In models where more than a single $D_3$-spurion appear, we
distinguish between the different $D_3$'s with an upper index, to
emphasize the fact that $d_3$ is different.

In terms of symmetries, the significance of $D_3$ is that it implies a
possible ${\cal O}(1)$ breaking of $SU(3)_Q\to SU(2)_Q \times U(1)_Q$
by the top Yukawa. The remaining symmetries are broken only by small
parameters and therefore constitute approximate symmetries in MFV
models. This is an important point that is valid in all single-Higgs
MFV models.\footnote{In multi-Higgs models at large $\tan \beta$, the bottom Yukawa could
provide an ${\cal O}(1)$ breaking of $SU(3)_D\to SU(2)_D \times
U(1)_D$.} We return to this point in the conclusions. 

Two comments are in order:
\begin{enumerate}
\item In models \textbf{QD} and \textbf{DD}, the $B_R$ and $D_R$
  fields transform in precisely the same way under both the gauge
  group and the global flavor group. We thus have freedom in choosing
  our basis in the $D_R-B_R$ space. We use this freedom to set
  $X_{BD}=0$.
\item Without fine-tuning, model \textbf{DQ} predicts non-hierarchical masses
  for the SM down quarks.  Two viable but fine-tuned solutions are
  to set $M_1=0$ or $m_2=0$.  We choose to work with the first,
  $M_1=0$. In Table \ref{tab:flarep} we denote a fined tuned value by
  a parenthesis.
\end{enumerate}

\section{Spectrum and couplings}
\label{sec:spe}
To understand the phenomenological aspects that are relevant to the
LHC, we have to find the spectrum and the couplings of the heavy
quarks.  Our starting point is the Lagrangian terms of
Eqs.~(\ref{eq:lagy}) and (\ref{eq:lagb}). We construct the down sector
mass matrices, diagonalize them, and obtain the spectrum of the heavy
and the light (i.e.~SM) quarks and the couplings of the heavy mass
eigenstates to the SM fields (a more detailed account of this
procedure will be given in subsequent work \cite{gntvz2}). We use
$B^\prime$ and $D^\prime$ to denote the heavy and the light down quark
mass eigenstates, respectively. We write the relevant couplings
schematically as follows:
\begin{eqnarray}
  \label{eq:4}
  {\cal L}_{B^\prime}=\overline{B_L^\prime}M_{B^\prime}B_R^\prime
  +\overline{D_L^\prime} Y_{B^\prime}^L B_R^\prime h
  +\overline{D_L^\prime} \gamma_\mu Y_{B^\prime}^T B_L^\prime Z^\mu
  +\overline{U_L^\prime} \gamma_\mu V_{\rm CKM} Y_{B^\prime}^T B_L^\prime W^\mu,
\end{eqnarray}
where $h$ is the physical Higgs field. $M_{B^\prime}$ is the diagonal
mass matrix of the heavy states.  In the $M_{B^\prime}\gg v$ limit,
the $B^\prime\to Z D^\prime$ and $B^\prime\to W U^\prime$ decays are
dominated by longitudinally polarized $Z$ and $W$ final states.
According to the Goldstone equivalence theorem, the sizes of the
corresponding decay rates are then given by $Y_{B^\prime}^L$ and
$V_{\rm CKM}Y_{B^\prime}^L$, respectively,\footnote{This is best seen
in the Feynman-t' Hooft gauge where the decays are predominantly
into unphysical Higgs states, with the relevant terms in the
Lagrangian $\overline{D_L^\prime} Y_{B^\prime}^L B_R^\prime h+
\overline{D_L^\prime} Y_{B^\prime}^L B_R^\prime h^3+
\overline{U_L^\prime}(\sqrt{2} V_{\rm CKM} Y_{B^\prime}^L)B_R^\prime
  h^+$.  See, for example, \cite{Perelstein:2003wd}.} with corrections
of order $M_W^2 / M_{B'}^2$.  The $Y_{B^\prime}^T$ matrix, on the
other hand, parametrizes the couplings of the transverse $W$ and $Z$
bosons.

If the $Y_UY_U^\dagger$ spurions could be neglected, then the flavor
structures would only depend on the CKM matrix $V_{\rm CKM}$ and the
diagonal down Yukawa coupling matrix $\hat{\lambda}$. Expressed in
approximate powers of the Wolfenstein parameter $\lambda\sim0.2$, we
have 
\begin{eqnarray}
  \label{eq:2}
  V_{\rm CKM}\sim\pmatrix{1&\lambda&\lambda^3\cr
    \lambda&1&\lambda^2\cr \lambda^3&\lambda^2&1\cr},\qquad
  \hat\lambda=\pmatrix{y_d&&\cr &y_s&\cr &&y_b\cr}\sim
  y_b\pmatrix{\lambda^4 &&\cr &\lambda^2&\cr &&1\cr}.
\end{eqnarray}
When the $Y_UY_U^\dagger$ effects are significant, the results are
modified in a simple way: the modification of the spectrum may involve
matrices of the form $D_3$, while the couplings may
involve a matrix $\tilde 1$:
\begin{eqnarray}
  \label{eq:1}
  \tilde 1 \equiv V_{\rm CKM}^\dagger D_3 V_{\rm CKM}\sim\pmatrix{
    1&0&\lambda^3\cr 0&1&\lambda^2\cr \lambda^3&\lambda^2&d_3\cr}, 
\end{eqnarray}
or matrices that scale in the same way with $\lambda$, for which we use the same symbol $\tilde 1$.

The masses and couplings for the various models are given in Table
\ref{tab:1}
with additional details of the derivation given in Appendix \ref{app-A}. We define a small parameter
\begin{eqnarray}
  \label{eq:3}
  \epsilon\equiv {v \over M},
\end{eqnarray}
where $v$ is the electroweak breaking scale, and $M \sim
\mbox{max}(M_1,M_2)$ is the heavy mass scale that fixes the masses of
the heavy quarks. Since the spectrum of the heavy quarks can be
hierarchical (models \textbf{QD} and \textbf{DQ}) or
(pseudo)degenerate (models \textbf{DD} and \textbf{QQ}), the heavy
mass scale $M$ differs significantly in the two cases. From the
requirement that the lightest $B'$ state has a mass in the TeV range,
one finds $\epsilon\sim 10^{-1}$ in models \textbf{DD} and
\textbf{QQ}, and $\epsilon\sim 10^{-5}$ in models \textbf{QD} and
\textbf{DQ}.

\begin{table}[t]
\begin{center}
\begin{tabular}{l|ccc} \hline\hline
\rule{0pt}{1.2em}%
~Model~ &\  $~~M_{B^\prime}/M~~$ &\ $~~Y_{B^\prime}^L~~$ &\
$~~Y_{B^\prime}^T~~$ 
 \\
\hline\hline
~~\textbf{QD} &\ $D_3\hat{\lambda}$ &$\tilde1\hat{\lambda} $ &\
$\epsilon\tilde1$\\ 
~~\textbf{DD} &\ 1&$ \tilde1\hat{\lambda} $ & $\epsilon
\tilde1\hat{\lambda} $\\ 
~~\textbf{DQ} &\ $D_3\hat{\lambda}$&$\tilde1$ & $\epsilon
\tilde1\hat{\lambda}^{-1}$\\ 
~~\textbf{QQ} &\ $D_3$&$\tilde1$ & $\epsilon \tilde1$\\ \hline\hline
\end{tabular}
\end{center}
\caption{The spectrum and couplings of the heavy quarks from
  Eq.~(\ref{eq:4}), given the flavor charges from Table
  \ref{tab:flarep}. $\hat{\lambda}$ is the diagonalized down Yukawa
  matrix, $\epsilon$ is the ratio of the electroweak scale to the
  heavy quark mass scale, and $\tilde{1} \equiv V_{\rm CKM}^\dagger
  D_3 V_{\rm CKM}$ parametrizes the effect of $SU(3)_Q$ breaking from
  the top Yukawa on the $B'$ couplings.}
\label{tab:1}
\end{table}

We learn the following points regarding the spectrum:
\begin{enumerate}
\item If the vector-like quarks are $SU(3)_Q$-singlets (model
  \textbf{DD}), the spectrum is fully degenerate. This degeneracy is
  lifted by effects of order $m_b^2/M^2$ that can be safely neglected.
\item If the vector-like quarks are $SU(3)_Q$-triplets (model
  \textbf{QQ}), the spectrum could have an interesting structure of
  $2+1$: two degenerate quarks and one with a mass of the same order
  of magnitude but not degenerate. This is a manifestation of the
  $O(1)$ breaking of $SU(3)_Q\to SU(2)_Q \times U(1)_Q$ due to $y_t$.
  The two degenerate states are split by effects of order $m_c^2/v^2
  \sim 10^{-4}$ that we neglect.
\item If the vector-like quarks are chiral (triplet+singlet) under
  $SU(3)_Q$ (model \textbf{QD} and \textbf{DQ}), the spectrum is
  hierarchical, with the hierarchy $y_d:y_s:{\cal O}(y_b)$.  In that
  case, only one heavy quark is at the TeV scale.
\end{enumerate}
As for the decay rates, we learn the following:
\begin{enumerate}
\item The decays to the transverse $W$ and $Z$ are always negligible,
  that is, $Y_{B^\prime}^T \ll Y_{B^\prime}^L$.
\item The couplings to longitudinal $W$/$Z$ and to $h$ are the
  same to a very good approximation. This implies that up to phase
  space effects, the heavy quarks decay rates to $W$, $Z$ and $h$ are
  in ratios $2:1:1$ \cite{Perelstein:2003wd}.
\item The flavor diagonal couplings dominate, that is
  $Y_{B^\prime}^{T,L}$ is close to a unit matrix.  The most
  significant flavor changing $Z$ coupling is
  $(Y_{B^\prime}^{L})_{23}\sim 0.04(Y_{B^\prime}^L)_{33}$ and the most
  significant flavor changing $W$ coupling is $(V_{\rm
    CKM}Y_{B^\prime}^L)_{12}\sim 0.23(V_{\rm
    CKM}Y_{B^\prime}^L)_{22}$.
\end{enumerate}

Finally, adding vector-like quarks to the SM affects, in general, the
low energy phenomenology of both flavor and electroweak precision
measurements. As concerns flavor, the CKM matrix is not unitary and
the $Z$-boson acquires flavor changing couplings to the down sector.
In the framework of MFV, the flavor changing $Z$ couplings are
suppressed by $\epsilon^2$, by small mixing angles and, in some
models, by down-sector Yukawa couplings. Consequently, these
contributions are safely within bounds. The effects of the extra
quarks on electroweak precision measurements are also suppressed by
$\epsilon^2$ \cite{Lavoura:1992np}. In some of the models, MFV leads
to further suppression of these effects \cite{gntvz2}. For $M\gsim$
TeV, the deviations of the $S$ and $T$ parameters from their SM values
are of ${\cal O}(0.01)$ in model \textbf{QQ}, and considerably smaller
in all other models. Thus, the models we study are generically allowed
by present data.

\section{LHC phenomenology}
\label{sec:lhc}
We are now ready to discuss the phenomenology of the model.  Our main
task is to check if the idea of MFV can be tested by direct
measurements at the LHC.  Clearly, we need to establish the fact that
new down-like quarks exist to start any probe of their flavor
structure.  An ATLAS study of vector-like down-type quarks using only
$2Z \rightarrow 4\ell$ final states found a $B'$ mass reach of 920 GeV
with $300 \mbox{ fb}^{-1}$ of data \cite{Mehdiyev:2006tz}, but the
inclusion of other $B'$ decay modes is likely to improve the reach,
given the small leptonic branching fraction of the $Z$.  For various
models with vector-like up-type quarks, the mass reach was found to
range from 1 to 2.5 TeV for $100-300 \mbox{ fb}^{-1}$ of data
\cite{Aguilar-Saavedra:2005pv, Skiba:2007fw, Azuelos:2004dm}.

The high end of the above discovery range is due to large mixing
angles with SM quarks, when the heavy quarks can be singly produced
using quark-$W$ fusion
\cite{Willenbrock:1986cr,Han:2003wu,Azuelos:2004dm}.  In our case,
such channels are particularly interesting for models \textbf{DQ} and
\textbf{QQ} where the couplings to longitudinal gauge bosons are
unsuppressed for the first generation, allowing the possibility for
$uW$ fusion to create a heavy $B_1'$.  Depending on the interplay
between parton distribution functions and flavor structures, the
single $B'$ channel may offer an interesting probe of minimal flavor
violation \cite{gntvz2}.

\begin{figure}
\includegraphics[scale=1.0]{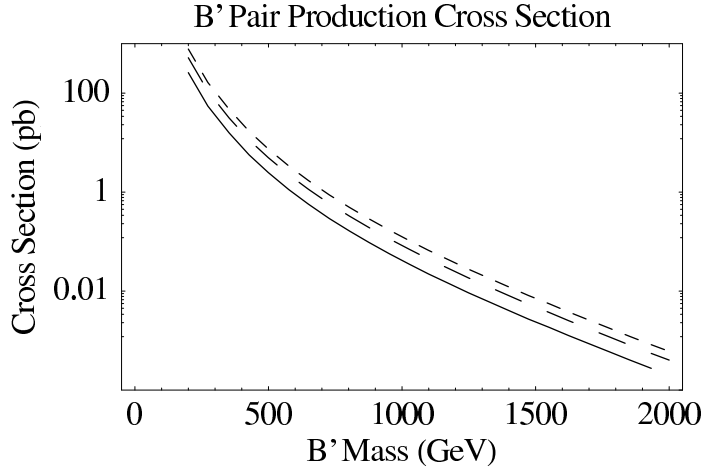}
\caption{\label{fig:xsec} Leading order cross section for $B'$ pair
  production at the LHC calculated at leading order using
  \texttt{Pythia 6.4.10} \cite{Sjostrand:2006za} with CTEQ5L parton
  distribution functions \cite{Lai:1999wy}.  From bottom to top, the
  total cross section for 1, 2, and 3 generations of $B'$ quarks. See
  \cite{Andre:2003wc} for the variation of the cross section from
  different choices  of factorization scale.}
\end{figure}

We focus on the QCD pair production channel $pp \rightarrow B'
\overline{B'}$ which is flavor diagonal by $SU(3)_C$ gauge invariance.
In Figure \ref{fig:xsec}, we show the estimated cross section for
$B^\prime$ pair production, calculated at leading order using
\texttt{Pythia 6.4.10} \cite{Sjostrand:2006za}.  After production,
each $B'$ quark decays to a SM quark and either a Higgs-, $Z$-, or
$W$-boson, leading to final states with multiple gauge bosons and hard
jets.

An important simplification of the analysis arises due to the absence
of missing energy involved with the new flavor physics.  Indeed by
assumption, the only new states are the heavy quarks, and except for
neutrinos from gauge boson decays, all final states can be observed.
Putting aside the question of backgrounds and signal efficiencies,
this would allow a determination of the $B'$ production cross sections
and the relative decay rates into $Wq$, $Zq$ and $hq$ (here $q$ stand
for any SM quark).\footnote{Depending on the Higgs mass and decay
  modes, this might be an interesting discovery channel for the Higgs.
  See, for example, \cite{Andre:2003wc}.}  With large enough
statistics, the $W$ and $Z$ helicities could be determined as well
\cite{who?}.

In order to separate $B'$ pair production from SM backgrounds, various
techniques can be used to reconstruct the $B'$ masses
\cite{Skiba:2007fw,Holdom:2007nw}.  Backgrounds for new vector-like
down-type quarks have also been studied in \cite{Andre:2003wc}.
Because we are interested in studying the flavor structure of $B'$
decays, though, we cannot rely too heavily on $b$-tagging to suppress
SM backgrounds.  On the other hand, unlike generic fourth generation
quarks, the $B'$ quarks have non-negligible branching fractions to
$Z$s, so requiring leptonic $Z$s can help suppress the large
$t\bar{t}$ and $W + \mbox{jets}$ backgrounds without biasing quark
flavor determination.

Though a complete background study is beyond the scope of the present
paper, example backgrounds calculated using \texttt{ALPGEN 2.11}
\cite{Mangano:2002ea} for a benchmark $B'$ mass of 600 GeV are shown
in Table \ref{table:background}.  Even in the most pessimistic case
where both a leptonic $Z$ and a leptonic $W$ are needed to reduce
backgrounds to an acceptable level, for three generations of $600$ GeV
$B'$ quarks, there can still be 2000 signal events at $100
\mbox{ fb}^{-1}$ with $O(1)$ signal to background ratio.\footnote{These
estimates make the unrealistic assumption that taus can be treated
on an equal footing with electrons and muons.  Given the large NLO
corrections to both QCD backgrounds and $B'$ pair production,
though, the estimate is still of the right order of magnitude.}

\begin{table}
\begin{tabular}{c||ccccc|cc|cc}
\hline
\hline
 & $~t \bar{t}~$ & $~t \bar{t} + j~$ & $~t \bar{t} + 2j~$ & $~W + 3j~$
 & $~W+4j~$ & $~Z + 3j~$ & $~Z + 4j~$ & $~WZ + 2j~$ & $~WZ + 3j~$ \\ 
 \hline
$~\sigma~$ & ~2.9 pb~ & 9.1 pb& 3.0 pb & (23.3 pb) & 4.4 pb & (2.0 pb)
& 0.5 pb & 0.020 pb& 0.006 pb\\
\hline
\hline
 & \multicolumn{5}{|c|}{$B'\overline{B'}$} &
 \multicolumn{2}{|c|}{$B'\overline{B'} \rightarrow ZX$} &
 \multicolumn{2}{|c}{$B'\overline{B'} \rightarrow WZX$} \\ 
\hline
$~\sigma~$ &  \multicolumn{5}{|c|}{2.7 pb} &
\multicolumn{2}{|c|}{0.14 pb} &  \multicolumn{2}{|c}{0.022 pb}\\
\hline
\hline
\end{tabular}

\caption{\label{table:background} Estimated SM backgrounds for three
generations of $B'$ pair production with a benchmark $B'$ mass of
600 GeV. These cross sections were calculated at leading order using
\texttt{ALPGEN 2.11} \cite{Mangano:2002ea} with CTEQ5L parton
distribution functions \cite{Lai:1999wy}. Jets have a minimum $p_T$ of
100 GeV with a $\Delta R$ separation of 1.0. For backgrounds involving
$W$s and $Z$s, the quoted cross section refers to gauge bosons
decaying to all three lepton generations (excluding $Z \rightarrow
\nu\nu$). To approximate the effect of cuts to isolate the $B'$ pair
production channel, the center-of-mass energy of the background events
are forced to be above $2 m_{B'}$. Backgrounds could be further
suppressed by insisting on $B'$ mass reconstruction
\cite{Skiba:2007fw,Holdom:2007nw}. The $W + 3j$ and $Z + 3j$
backgrounds are in parentheses because they are only backgrounds in
regions of phase space where the $W/Z/h$ from a $B'$ decay is
boosted enough to form one ``fat'' jet. The ``$B'\overline{B'}
\rightarrow ZX$'' cross section assumes that the $W$:$Z$:$h$ branching
ratios are in a $2$:$1$:$1$ ratio and the $Z$ decays to visible
leptons (including taus). The ``$B'\overline{B'} \rightarrow WZX$''
cross section requires an additional leptonic $W$.}
\end{table}

To test MFV, one must extract information about the spectrum of the
heavy quarks and their partial and total decay widths.  Especially
important are the tagging capabilities of the LHC.  The quark from the
$B'$ decay can be classified either as a light jet, a heavy-flavor
($b$ or $c$) jet, or a $t$ quark (by identifying $bW$ combinations
with the right invariant mass).  The purities and efficiencies depend
on the flavor, the energy of the jet, and the details of the rest of
the event.\footnote{Combinatoric background in $t$ reconstruction is
  an obvious challenge for high-multiplicity final states.  The large
  boost factor of the $B'$ decay products may alleviate some of the
  combinatoric issues, though.}  We expect that the ability to
distinguish a $t$ quark from a $c$ or $u$ quark will be a particularly
good handle because of uncertainties in the heavy-flavor tagging
efficiencies at high $p_T$.  That is, for heavy-flavor and light jets
alone, if the flavor violation is not large, it can perhaps 
be reinterpreted as flavor
conservation with modified heavy-flavor efficiencies.  Top quarks are
special because their decays add additional $W$s to events, making it
in principle easier to tag the third generation in a robust way.  Of
course, QCD radiation can easily add $80 \mbox{ GeV}$ of energy to an
event, so the ability to ``$t$-tag'' may depend on the ability to
simultaneously reconstruct the $B'$ and top masses.  A more detailed
study of these issues is needed in order to see how ambitious this
flavor program can become \cite{gntvz2}.

In what follows, we investigate what can be learned about MFV if the
LHC provides us with (i) the spectrum, (ii) some capability for heavy
flavor tagging, and (iii) some information on the decay widths.

(i) {\it Spectrum}. MFV predicts that at the TeV scale, there is
either a near degenerate spectrum of $B'$ quarks, or only one
kinematically accessible flavor.  A discovery of two (or more)
non-degenerate states at the TeV scale will disfavor MFV. (MFV will not be excluded because the two
 non-degenerate states might be the lightest members of two
 different triplets.) Conversely,
by measuring the mass and the production cross section, the LHC might
be able to establish that there is a three-fold or two-fold
degeneracy.  That will provide evidence that some flavor $SU(3)$ or
$SU(2)$ is at work.

In principle, the total cross section can tell us the degeneracy of
each state by comparing the latter to the $t\bar t$ production rate.
The extraction of the degeneracy requires theoretical knowledge of the
mass dependence of the production cross section, and experimental
measurements of the mass of the heavy quarks, their production cross
section, and the $t\bar t$ production rate. A complication in this procedure comes from the
different decay modes of the $B^\prime$ compared to the $t$. It would
be helpful to measure several decay modes of the $B^\prime$ to confirm
the expected $W/Z/h$ branching ratios. If it can be determined that
$B^\prime$ decays always involve longitudinally polarized $W$s and $Z$s,
then this could be used as a further argument for why the $W/Z/h$
branching ratios should be fixed by the Goldstone equivalence theorem.

A three-fold degeneracy might get further support by looking at the
flavor content of $B'$ pair production events. Since the $B'$ quarks
are produced in pairs, MFV predicts that 1/3 of the pairs decay
exclusively into third generation quarks, while 2/3 into non-third
generation quarks.  Such evidence will make the case for degeneracy
strong and will provide a rather convincing evidence for MFV.

In cases that the $B^\prime$ quarks are too heavy to be pair-produced
in a statistically significant amount, the single $B'$ production
can still be significant. This is particularly true for models
\textbf{DQ} and \textbf{QQ}, where $(Y_{B^\prime}^L)_{11}={\cal
  O}(1)$. Here, the single $B^\prime$ production channel has the
peculiar feature that the production rates are determined by parton
distribution functions. Furthermore, it can be used to test MFV,
because the singly produced $B_1^\prime$ should not decay to third
generation quarks.    

(ii) {\it Flavor tagging}. The hierarchy $v\ll M$ guarantees that the
rates into the three different final bosons are comparable,
\begin{eqnarray}
  \label{eq:5}
  \Gamma(B'\to Wq)\approx 
  2\Gamma(B'\to Zq)\approx 
  2\Gamma(B'\to hq).
\end{eqnarray}
Thus, the LHC can use whichever (or a combination) of these modes that
is optimal for flavor tagging.  As mentioned above, because of the
large $t\bar{t}$ and $W + \mbox{jets}$ backgrounds, events with at
least one leptonically decaying $Z$ are likely to be the most useful.

The most prominent feature of the MFV models is the suppression of
flavor changing couplings: each mass eigenstate decays to a very good
approximation only to SM quarks of the corresponding generation. This
property is a direct consequence of MFV. Namely, all flavor violating effects
are proportional to the CKM matrix, which is very close to the unit
matrix. It is this feature of MFV that can be tested in the decays of
the heavy quarks.

Flavor tagging will therefore allow the LHC to put MFV to the test.
First, consider events where the heavy quarks are pair produced. MFV
predicts that both of them should decay to quarks of the same
generation. Since the mixing between the third generation to the
light one is of order $|V_{cb}|\sim 0.04$, we can test the following
prediction:
\begin{eqnarray}
  \label{eq:6}
  \frac{\Gamma(B'\overline{B'} \to X q_{1,2} q_3)}
  {\Gamma(B'\overline{B'} \to X q_{1,2} q_{1,2})+\Gamma(B'\overline{B'} \to X q_3 q_3)} 
  \lsim 10^{-3}.
\end{eqnarray}
Here $q_3$ stands for third generation quarks ($b,t$), $q_{1,2}$
stands for first two generation quarks ($u,d,s,c$) and both $q_3$ and
$q_{1,2}$ stand for both quarks and antiquarks. Note that Eq.~(\ref{eq:6}) is a non-trivial check of MFV, because constraints from low energy flavor experiments \cite{Yanir:2002cq} 
still allow flavor-changing couplings in $Y_{B^\prime}^L$ of Eq. 
(\ref{eq:4}) that are considerably larger than those predicted by
MFV.  In fact, this ratio could even be ${\cal O}(1)$.

Second, in the case that there is no degeneracy at all, MFV predicts
that each mass eigenstate decays either to third generation quarks or
to light quarks, to an accuracy of $O(10^{-3})$.  In the case of
twofold degeneracy, MFV predicts that the two mass eigenstates decay
to light quarks only, up to ${\cal O}(10^{-3})$ effects.

Finally, if charm tagging is also possible, the theory can be tested
further.  Consider a non-degenerate state that decays into light
quarks (for example, model \textbf{QD}). MFV implies that this light state must
decay predominantly to the first generation with small charm branching
ratio, of order $\lambda^2\sim5\%$. A larger amount of
charm will therefore exclude MFV.

(iii) {\it Decay width}.  In principle, measurements of the total
decay widths of degenerate states can provide a smoking gun signal
since their ratio is either one to a good accuracy (model \textbf{QQ})
or is given by the ratio of light quark masses (model \textbf{DD}).
Unfortunately, it seems unlikely that the total decay width of the
states can be measured. In models \textbf{QD} and \textbf{DD}, the
width is, on one hand, highly suppressed and far below the
experimental resolution, and on the other hand, much larger than the
width required to generate a secondary vertex.\footnote{There is the
amusing possibility in models \textbf{QD} and \textbf{DD} of
fine-tuning the overall magnitude of the $Y^L_{B'}$ coupling to be
small while still maintaining MFV, allowing the $B'_1$ to be
long-lived enough to generate a secondary vertex while the $B'_3$
decays promptly.} In models \textbf{DQ} and \textbf{QQ}, the width
is roughly of the size of the experimental resolution ($3\%$), which
gives hope that we may get some information on the width. 

As a final remark, we note that perhaps the most spectacular case will
arise if model \textbf{QQ} is realized in Nature, with all three heavy
quarks within reach of the LHC. Establishing a $2+1$ spectrum, with
the separated quark decaying exclusively into third generation quarks,
and the two degenerate states decaying exclusively into non-third
generation quarks will provide convincing evidence for MFV.  In fact,
a two-fold degeneracy which involves no third generation quarks will
probably be sufficient to support MFV.

\section{Conclusions}
\label{sec:con}
We have explored the question of whether high $p_T$ physics at the LHC
can contribute to our understanding of flavor physics. We considered
here a specific framework of new physics, that of extra down-type
SU(2)-singlet quarks in the simplest representations under the flavor
group. Many other possibilities can be considered \cite{gntvz2}: new
down-like quarks in other representations of the flavor group, such as
triplets of $SU(3)_U$; up-type SU(2)-singlet quarks; extra weak
doublets; or even extra heavy leptons \cite{Cirigliano:2004mv}.

Our scenario spans, however, four representative situations: the
spectrum can be degenerate or hierarchical, and the couplings to SM
quarks can be universal or hierarchical. Our framework demonstrates
that, in spite of this variety of options, there are several features
that are common to all MFV models. 

In particular, our main result, that extra quarks at the TeV scale
will allow the LHC to test MFV, does not depend on the specific
implementation of MFV. MFV implies that the new physics is, to a very
good approximation, flavor conserving. Thus, by roughly testing the
flavor structure of the new quarks, MFV can, in principle, be excluded
or, otherwise, supported and probed.

The more detailed structure of the MFV principle can be tested in
various ways. The full symmetry in the down sector is $SU(3)_Q\times
SU(3)_D$. In model \textbf{DD}, one can achieve evidence for this
symmetry from the threefold degeneracy. The only order one breaking of
the flavor symmetry in the down sector is due to $Y_UY_U^\dagger$. It
breaks $SU(3)_Q\times SU(3)_D\to SU(2)_Q\times U(1)_Q\times SU(3)_D$.
In model \textbf{QQ}, one can see evidence for this breaking by
observing a $2+1$ spectrum. Further evidence for the approximate
symmetry can be obtained in all models from the decays of heavy quarks
which do not mix third generation with first and second. The down
quark masses $\hat\lambda$ lead to further breaking into $U(1)_b\times
U(1)_s\times U(1)_d$. Measuring this breaking requires sufficient
$c$-tagging (which can perhaps be achieved). The effects of
$U(1)_s\times U(1)_d$ breaking are proportional to $|V_{us}|^2$;
measuring them via the small rate of $B^\prime\overline{B^\prime}
\rightarrow ZdWc$ will be very hard at the LHC without excellent
$c$-tagging efficiency. The $U(1)_b$ breaking effects are proportional
to $|V_{cb}|^2$ and therefore below the observable level.
Consequently, they provide the strongest test of MFV.

Going forward, the main experimental issues that must be understood
with regard to high-$p_T$ flavor studies are:
\begin{itemize}
\item How well will the heavy-flavor tagging efficiency be known at
  high-$p_T$?  Because flavor-violation could be masked by adjustments
  in the $b$-tagging efficiency, it may be desirable to develop less
  efficient but better calibrated $b$-tagging methods.
\item What are the prospects for ``$t$-tagging'' in high multiplicity
  events?  The ability to robustly identify when events have extra
  $W$s from top decays will aid in the identification of $B'$ decays
  to the third generation.
\item Assuming the $B'$ mass is measured in a clean channel, to what
  extent is it possible to separate SM backgrounds from $B'$ signals
  using $B'$ mass reconstruction?  Because flavor studies are likely
  to be statistics limited, it may be desirable to use events with
  fewer numbers of final state leptons, for which $t\bar{t}$ and $W/Z
  + \mbox{jets}$ backgrounds are substantial.
\end{itemize}

We conclude that if the LHC discovers new particles, it can also make
a significant contribution to our understanding of flavor physics.
The confirmation or invalidation of the MFV hypothesis will illuminate
the new physics flavor puzzles, providing insight into the relation
between high precision tests at low energy and new discoveries at the
energy frontier.

\textbf{Acknowledgements:}  We thank Roni Harnik, Zoltan Ligeti,
Michelangelo Mangano, Gilad Perez, and Yael Shadmi for helpful
discussions. This project was supported by the Albert Einstein Minerva
Center for Theoretical Physics. The work of Y.G. is supported in part
by the Israel Science Foundation under Grant No.~378/05.
The research of Y.N. is supported by
the Israel Science Foundation founded by the Israel Academy of
Sciences and Humanities, the United States-Israel Binational Science
Foundation (BSF), Jerusalem, Israel, the German-Israeli foundation for
scientific research and development (GIF), and the Minerva Foundation.
The work of J.T. is supported by a fellowship from the Miller
Institute for Basic Research in Science. The work of J.Z. is supported in 
part by the European Commission RTN network, Contract No.~MRTN-CT-2006-035482 
(FLAVIAnet) and by the Slovenian Research Agency. 

\appendix
\section{Calculation of spectra and couplings}\label{app-A}
In this appendix, we derive the heavy quark spectra that follow from the mass terms in Eqs.~(\ref{eq:lagy}) and (\ref{eq:lagb}).
After electroweak symmetry breaking, these terms take the following form:
\beq\label{A1}
{\cal L}_m=\pmatrix{\overline{Q}_L \cr \overline{B}_L }^{\mathrm T} \underbrace{\pmatrix{v Y_D & m_2 Y_B \cr M_1 X_{BD} & M_2 X_{BB} \cr}}_{M_d} \pmatrix{D_R \cr B_R \cr},
\eeq
with the transpose acting only on the generation indices. The values of the matrices $Y_B , X_{BD} , X_{BB}$ for the four models are compiled in Table \ref{tab:flarep}, while the resulting spectra are given in Table \ref{tab:1}. These results are easiest to derive by diagonalizing  $M_d M_d^\dagger$, where $M_d$ is the mass matrix in Eq.~(\ref{A1}).
Taking $v\sim m_2\ll M_{1,2}$ the matrix $M_d M_d^{\dagger}$ scales as (suppressing the dependence on $Y_D$ and $D_3$)
\beq
M_d M_d^{\dagger}\sim \pmatrix{v^2 & v M \cr v M & M^2 \cr}.
\eeq
Up to $v/M$ suppressed corrections, the mass spectrum of heavy states is thus governed by the lower right $3\times 3$ submatrix
\beq\label{MMdagger}
(M_d M_d^\dagger)_{BB}=M_1^2 X_{BD} X_{BD}^\dagger+M_2^2 X_{BB} X_{BB}^\dagger\simeq M_2^2 X_{BB} X_{BB}^\dagger,
\eeq
where the last equality is valid up to Yukawa suppressed terms. For the {\bf DD} and {\bf QQ} models this is already diagonal, giving spectra as in Table \ref{tab:1}.
For the {\bf QD} and {\bf DQ} models it is useful to go to the basis where $Y_D$ is diagonal. The diagonalization is done by the biunitary transformation
\beq\label{VLR}
V_L Y_D V_R^\dagger=Y_D^{\rm diag}\equiv\hat\lambda.
\eeq
In this basis, the spectra in Table \ref{tab:1} then follow immediately, if one neglects the remaining small off-diagonal terms and uses the relation $V_{\rm CKM}=V_L^\dagger+O(v^2/M^2)$. Explicitly, for the ${\bf QD}$ model we have
\beq
X_{BB}X_{BB}^\dagger\to V_L X_{BB}X_{BB}^\dagger V_L^\dagger=V_L D_3 Y_D Y_D^\dagger D_3 V_L^\dagger=V_L D_3 V_L^\dagger \hat \lambda^2 V_L D_3 V_L^\dagger,
\eeq
which together with Eq.~(\ref{eq:1}) gives $M_{B'}\sim M D_3 \hat \lambda$ as in Table \ref{tab:1}.

In a similar way, the spectrum could be obtained from $M_d^\dagger M_d$. Note that the  $3\times 3$ off-diagonal blocks are either $v^2/M^2$ or $Y_D$ suppressed (the large right-handed rotations between $D_R$ and $B_R$ in ${\bf QD}$ and ${\bf DD}$ models have already been used to set $X_{BD}=0$). As above, the heavy quark spectra then follow from
\beq
(M_d^\dagger M_d)_{BB}=m_2^2 Y_B^\dagger Y_B+M_2^2 X_{BB} X_{BB}^\dagger\sim M_2^2 X_{BB}^\dagger X_{BB},
\eeq
giving the same result as Eq.~(\ref{MMdagger}). The diagonalization of $(M_d^\dagger M_d)_{BB}$ also gives the Higgs coupling
\beq \label{Higgscoupl}
\frac{m_2}{v}\overline{Q_L}Y_B B_R H,
\eeq
in the $Q_L'$, $B_R'$ mass-eigenstate basis. Up to suppressed terms we have
\beq\label{rotations}
Q_L=U_L^\dagger Q_L', \qquad B_{R}=U_{R}^\dagger B_{R}',
\eeq
where $U_{R}$ diagonalizes the $(M_d^\dagger M_d)_{BB}$ matrix and $U_L\sim V_{\rm CKM}$. The similarity sign means that the two sides scale in the same way in terms of the Wolfenstein parameter $\lambda$, with the scaling for $V_L=V_{CKM}^\dagger +O(v^2/M^2)$ given in Eq.~(\ref{eq:2}). In the mass eigenstate basis, the Higgs coupling is
\beq \label{Higgscouplmass}
\frac{m_2}{v}\overline{Q_L'}U_L Y_B U_R^\dagger B_R H.
\eeq
From this, the scaling of Higgs couplings with the Wolfenstein parameter $\lambda$  is obtained as given in Table \ref{tab:1}. In the derivation one also uses the fact that $Y_UY_U^\dagger$ insertions preserve an $SU(2)$ symmetry of the first two generations, so that the Higgs coupling $U_L Y_B U_R^\dagger$ is diagonal in the first two rows and columns.




\begin{thebibliography}{99}
\bibitem{Yao:2006px}
  W.~M.~Yao {\it et al.}  [Particle Data Group],
  J.\ Phys.\ G {\bf 33}, 1 (2006).

\bibitem{Group:2007cr}
  H.~F.~A.~Group,
  arXiv:0704.3575 [hep-ex].

\bibitem{Hinchliffe:2000np}
  I.~Hinchliffe and F.~E.~Paige,
  Phys.\ Rev.\  D {\bf 63}, 115006 (2001)
  [arXiv:hep-ph/0010086];
  D.~F.~Carvalho, J.~R.~Ellis, M.~E.~Gomez, S.~Lola and J.~C.~Romao,
  Phys.\ Lett.\  B {\bf 618}, 162 (2005)
  [arXiv:hep-ph/0206148];
  K.~Agashe and M.~Graesser,
  Phys.\ Rev.\  D {\bf 61}, 075008 (2000)
  [arXiv:hep-ph/9904422];
  N.~V.~Krasnikov,
  JETP Lett.\  {\bf 65}, 148 (1997)
  [arXiv:hep-ph/9611282];
  A.~Bartl, K.~Hidaka, K.~Hohenwarter-Sodek, T.~Kernreiter, W.~Majerotto and W.~Porod,
  Eur.\ Phys.\ J.\  C {\bf 46}, 783 (2006)
  [arXiv:hep-ph/0510074].

\bibitem{D'Ambrosio:2002ex}
  G.~D'Ambrosio, G.~F.~Giudice, G.~Isidori and A.~Strumia,
  Nucl.\ Phys.\  B {\bf 645}, 155 (2002)
  [arXiv:hep-ph/0207036].
  
\bibitem{Buras:2000dm}
  A.~J.~Buras, P.~Gambino, M.~Gorbahn, S.~Jager and L.~Silvestrini,
  Phys.\ Lett.\  B {\bf 500}, 161 (2001)
  [arXiv:hep-ph/0007085].

\bibitem{Buras:2003jf}
  A.~J.~Buras,
  Acta Phys.\ Polon.\  B {\bf 34}, 5615 (2003)
  [arXiv:hep-ph/0310208].

\bibitem{Branco:1999fs}
  G.~C.~Branco, L.~Lavoura and J.~P.~Silva,
  ``CP violation,''
{\it  Oxford, UK: Clarendon (1999)}.

\bibitem{Aguilar-Saavedra:2002kr}
  J.~A.~Aguilar-Saavedra,
  Phys.\ Rev.\  D {\bf 67}, 035003 (2003)
  [Erratum-ibid.\  D {\bf 69}, 099901 (2004)]
  [arXiv:hep-ph/0210112].

\bibitem{Andre:2003wc}
  T.~C.~Andre and J.~L.~Rosner,
  Phys.\ Rev.\  D {\bf 69}, 035009 (2004)
  [arXiv:hep-ph/0309254].

\bibitem{Yanir:2002cq}
  T.~Yanir,
  JHEP {\bf 0206}, 044 (2002)
  [arXiv:hep-ph/0205073];
  G.~Barenboim, F.~J.~Botella and O.~Vives,
  Nucl.\ Phys.\  B {\bf 613}, 285 (2001)
  [arXiv:hep-ph/0105306];
  T.~Morozumi, Z.~H.~Xiong and T.~Yoshikawa,
  arXiv:hep-ph/0408297;
  J.~A.~Aguilar-Saavedra,
  Phys.\ Rev.\  D {\bf 67}, 035003 (2003)
  [Erratum-ibid.\  D {\bf 69}, 099901 (2004)]
  [arXiv:hep-ph/0210112].

\bibitem{gntvz2}
Y. Grossman, Y. Nir, J. Thaler, T. Volansky, and J. Zupan, work in progress.

\bibitem{Perelstein:2003wd}
  M.~Perelstein, M.~E.~Peskin and A.~Pierce,
  Phys.\ Rev.\  D {\bf 69}, 075002 (2004)
  [arXiv:hep-ph/0310039].

\bibitem{Lavoura:1992np}
  L.~Lavoura and J.~P.~Silva,
  Phys.\ Rev.\  D {\bf 47}, 2046 (1993).

\bibitem{Mehdiyev:2006tz}
  R.~Mehdiyev, S.~Sultansoy, G.~Unel and M.~Yilmaz,
  Eur.\ Phys.\ J.\  C {\bf 49}, 613 (2007)
  [arXiv:hep-ex/0603005].

\bibitem{Aguilar-Saavedra:2005pv}
  J.~A.~Aguilar-Saavedra,
  Phys.\ Lett.\  B {\bf 625}, 234 (2005)
  [Erratum-ibid.\  B {\bf 633}, 792 (2006)]
  [arXiv:hep-ph/0506187].

\bibitem{Skiba:2007fw}
  W.~Skiba and D.~Tucker-Smith,
  arXiv:hep-ph/0701247.

\bibitem{Azuelos:2004dm}
  G.~Azuelos {\it et al.},
  Eur.\ Phys.\ J.\  C {\bf 39S2}, 13 (2005)
  [arXiv:hep-ph/0402037].
  
\bibitem{Willenbrock:1986cr}
  S.~S.~D.~Willenbrock and D.~A.~Dicus,
  Phys.\ Rev.\  D {\bf 34}, 155 (1986).

\bibitem{Han:2003wu}
  T.~Han, H.~E.~Logan, B.~McElrath and L.~T.~Wang,
  Phys.\ Rev.\  D {\bf 67}, 095004 (2003)
  [arXiv:hep-ph/0301040].

\bibitem{Sjostrand:2006za}
  T.~Sjostrand, S.~Mrenna and P.~Skands,
  JHEP {\bf 0605}, 026 (2006)
  [arXiv:hep-ph/0603175].

\bibitem{Lai:1999wy}
  H.~L.~Lai {\it et al.}  [CTEQ Collaboration],
  Eur.\ Phys.\ J.\  C {\bf 12}, 375 (2000)
  [arXiv:hep-ph/9903282].

\bibitem{who?}
  F.~Hubaut, E.~Monnier, P.~Pralavorio, K.~Smolek and V.~Simak,
  Eur.\ Phys.\ J.\  C {\bf 44S2}, 13 (2005)
  [arXiv:hep-ex/0508061].

\bibitem{Holdom:2007nw}
  B.~Holdom,
  JHEP {\bf 0703}, 063 (2007)
  [arXiv:hep-ph/0702037];
  arXiv:0705.1736 [hep-ph].
  
\bibitem{Mangano:2002ea}
  M.~L.~Mangano, M.~Moretti, F.~Piccinini, R.~Pittau and A.~D.~Polosa,
  JHEP {\bf 0307}, 001 (2003)
  [arXiv:hep-ph/0206293].

\bibitem{Cirigliano:2004mv}
  V.~Cirigliano, A.~Kurylov, M.~J.~Ramsey-Musolf and P.~Vogel,
  Phys.\ Rev.\  D {\bf 70}, 075007 (2004)
  [arXiv:hep-ph/0404233].

\end{thebibliography}
\end{document}